\documentclass[draftcls,12pt,onecolumn,oneside]{IEEEtran}
\usepackage{mathrsfs}
\usepackage{url,times,amsmath,color,amssymb,graphicx,epsfig,psfig,cite,geometry,psfrag,subfigure,dsfont,booktabs}
\usepackage{algorithm}
\usepackage{algorithmic}
\usepackage{bm}
\begin{document}
\title{Fog Radio Access Networks: Mobility Management, Interference Mitigation and Resource Optimization}
\author{Haijun Zhang,~\IEEEmembership{Senior Member,~IEEE}, Yu Qiu, Xiaoli Chu,~\IEEEmembership{Senior Member,~IEEE}, \\
  Keping Long,~\IEEEmembership{Senior Member,~IEEE}, and Victor C.M. Leung,~\IEEEmembership{Fellow,~IEEE}
\thanks{Haijun Zhang and  Keping Long are with the Beijing Engineering and Technology Research Center for Convergence Networks, University of Science and Technology Beijing, Beijing, 100083, China (e-mail: haijunzhang@ieee.org, longkeping@ustb.edu.cn).

Yu Qiu is with College of Information Science and Technology, Beijing University of Chemical Technology, Beijing, 100029, China (Email: eeqiuyu@gmail.com).

Xiaoli Chu is with Department of Electronic and Electrical Engineering, the University of Sheffield, Sheffield S1 3JD, UK (Email: x.chu@sheffield.ac.uk).

Victor C.M. Leung is with the Department of Electrical and Computer Engineering, The University of British Columbia, Vancouver, BC V6T 1Z4 Canada (e-mail: vleung@ece.ubc.ca).

This work was supported by the National Natural Science Foundation of China (Grant 61471025), the Young Elite Scientist Sponsorship Program by CAST (2016QNRC001), and the Chinese Fundamental Research Funds for the Central Universities.

}}
\maketitle

\begin{abstract}
In order to make Internet connections ubiquitous and autonomous in our daily lives, maximizing the
utilization of radio resources and social information is one of the major research topics in future mobile
communication technologies. Fog radio access network (FRAN) is regarded as a promising paradigm
for the fifth generation (5G) of mobile networks. FRAN integrates fog computing with RAN and makes
full use of the edge of networks. FRAN would be different in networking, computing, storage and
control as compared with conventional radio access networks (RAN) and the emerging cloud RAN.
In this article, we provide a description of the FRAN architecture, and discuss how the distinctive
characteristics of FRAN make it possible to efficiently alleviate the burden on the fronthaul, backhaul
and backbone networks, as well as reduce content delivery latencies. We will focus on the mobility
management, interference mitigation, and resource optimization in FRAN. Our simulation results show
that the proposed FRAN architecture and the associated mobility and resource management mechanisms
can reduce the signaling cost and increase the net utility for the RAN.
\end{abstract}

\section{Introduction}

Nowadays billions of mobile users receive seamless and stable wireless services supported
by the communication infrastructures. With the rapid development of mobile communications,
dozens of network standards have emerged, including the Third Generation Partnership Project
(3GPP) Long Term Evolution (LTE) standards. In addition to advanced multiple-input multipleoutput
(MIMO) technologies \cite{MD2011}, small cells and heterogeneous networks \cite{AD2011}, cloud radio access
networks (CRAN) have emerged as a popular technology for future mobile networks \cite{YL2010}. CRAN
features centralized resource management, with all the computing, control, and data storage of
the network gathered into the cloud. The centralized data centers, cellular core networks and backbone networks are equipped with computing, storage and network management functions.
However, recent research \cite{AD2013, HCRANmugenIWC2014} has shown that the completely centralized architecture of CRAN makes it hard to cope with the unpredictable mobility of users, the increasing density of base
stations (BSs) \cite{NB2014}, and the explosive growth of user data demand. The planning and optimization
of heterogeneous networks are facing complicated inter-cell interference problems and increased
management complexities. More recently, heterogeneous cloud radio access network (HCRAN)
has been proposed \cite{SH2015}, where remote radio heads (RRHs) working in coordination with high
power nodes can effectively mitigate co-channel interference. Although HCRAN may offer better
cost efficiency than CRAN \cite{SH2015}, the complex cost structure behind HCRAN, how its resource
optimization should be supported by the baseband unit (BBU) pool, and its traffic burden on the
cloud center require more in-depth studies. Since all information is exchanged through the BBU
pool, HCRAN may cause additional burden on the fronthaul and backhaul links, especially
the wireless ones \cite{DCC2014}, as compared with CRAN. In the meanwhile, more data is generated
from various social media platforms due to their increasing popularity. Hence, it becomes
increasingly important to consider social networking and local information in the management
and optimization of RANs. This is not easily achieved in CRAN or HCRAN because of their
centralized architecture. 

In view of the above issues related to CRAN and HCRAN and the requirements of the future communication scenarios, FRAN was introduced by Cisco to exploit local signal processing and computing, cooperative resource management,
and distributed storing/caching capabilities at the network edge \cite{FB2012}. In FRAN, a large amount
of signal processing and computing is performed in a distributed manner, rather than all by
the centralized BBU pool, and local data can be stored in edge devices, such as access points
(APs) and user equipment (UE), instead of the cloud data center. A unique feature of FRAN
is to maximize the use of edge devices of the network, e.g., to perform collaborated radio
signal processing. As a result, the burden on the fronthaul is much relieved than CRAN or
HCRAN. Due to these distinctive characteristics of FRAN, network management and optimization
mechanisms need to be revisited for FRAN. In \cite{Haijun2015}, cooperative interference mitigation and
handover management were studied for heterogeneous cloud small cell networks. An information-centric
wireless network virtualization framework was studied in \cite{CLF2016}. These requirements translate into a tremendous demand for bandwidth and energy consumption.
MIMO is a promising solution for sloving these
issues as it provides extra degrees of freedom in the spatial domain which promote a tradeoff
between diversity gain and multiplexing gain. Over the past few year, an enormous
amount of research has been concentrated on MIMO communication \cite{FZ2014}. However, the
modest computational capabilities of mobile devices limit the MIMO gains that can be achieved in practice. An attractive alternative for realizing the performance gains offered by
multiple antennas is multiuser MIMO, where a multipleantenna transmitter serves multiple
single-antenna receivers simultaneously. In fact, the combination of multiuser MIMO and
distributed antennas is widely recognized as a viable technology for extending service
coverage and mitigating interference \cite{RH2013}. Specifically, distributed antennas introduce additional
capabilities for combating both path shadowing and loss by shortening the distances
between the the receivers and transmitters. Furthermore, even with these powerful MIMO
techniques, spectrum scarcity is still a main obstacle in providing high speed uplink and
downlink communications. In this article, we only consider the signal antenna system and will consider the MIMO beamforming
based interference management in our future work. 
However, mobility management, interference mitigation, and resource optimization for FRAN have not been sufficiently studied in the existing works.

In this article, we first present a network architecture of FRAN. Then, we propose the handover
management and handover procedures, which make use of edge caching, in FRAN. For effective
interference and resource management, we introduce an interference-aware price and a price
of using the fronthaul for caching in the network edge. The subchannel and power allocation
problem is then modeled as a non-cooperative game to optimize the resource allocation in
FRAN. Moreover, we analyze the signaling overhead in FRAN using numerical results. We will
show that the proposed interference and resource management mechanisms not only reduce the
interference in the FRAN, but also enhance the net utility of the network.

\section{System Architecture}

Fig. 1 shows the FRAN architecture, where the marco remote radio heads (MRRHs), small
RRHs (SRRHs) and fog-computing access points (F-APs) all connect to the BBU pool. We denote
the smart user equipment as F-UE in the FRAN. The MRRHs are connected to the BBU pool by
the backhaul links. The SRRHs and the F-APs are connected to the BBU pool by the fronthaul
links. The F-APs, which are unique to FRANs, integrate not only the fronthaul radio frequency
but also the physical-layer signal processing functionalities and procedures of the upper layers.
Thus, F-APs can implement collaborative radio signal processing locally using their adequate
computing capabilities and can manage their caching memories flexibly. It is worth noting that
although both SRRHs and F-APs are equipped with caching capabilities, the contents stored
in them are fundamentally different: the contents cached in F-APs are highly locally popular
or relevant, but not those cached in SRRHs. With the increasing popularity of location-based
mobile applications, a lot of superfluous information may be generated adding to the surging data
traffic over the fronthaul between the SRRHs and the centralized BBU pool, which pushes the
fronthaul links to their capacity limits. In fact, the social application data exchanged between
neighboring F-UEs shows a high degree of conformity. Some social applications would only
generate data traffic between F-UEs in close physical proximity. Besides, users from the same social group or having the same social interest may request the same contents over the downlink.
In these cases, F-APs can provide the requested services locally by caching the popular contents.
Consequently, users do not need to connect to the BBU pool every time when they require
data or contents.

\begin{figure}[t]
        \centering
        \includegraphics*[width=13cm]{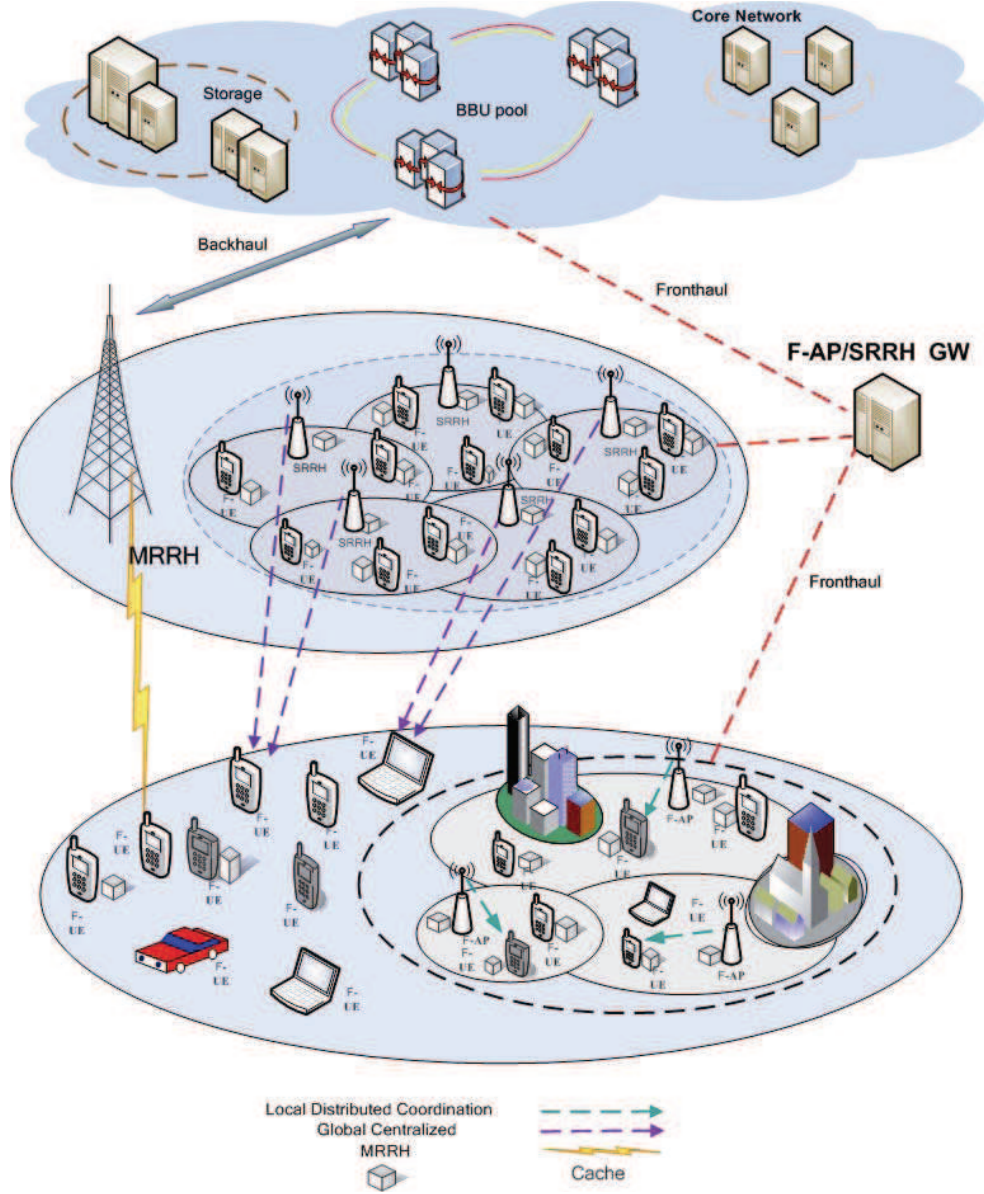}
        \caption{An FRAN architecture.}
        \label{fig:1}
\end{figure}

High mobility F-UEs are served by MRRHs. Although the interface between
the BBU pool and the MRRHs on the backhaul link is compatible with that defined in the
3GPP standards for LTE and LTE-Advanced systems \cite{MP2016}, the BBU pool will mainly provide
centralized storage and communications in FRAN. SRRHs connect to the BBU pool in order to
retrieve the packets cached at the BBU pool or packets from the cloud network. SRRHs are also
equipped with caching capabilities to provide local caching services to their associated F-UE.
Collaborative signal processing can be performed by SRRHs, F-APs and F-UEs locally, and
cooperative resource management is performed by F-APs and F-UEs in a distributed manner.
Consequently, the communications burden on the fronthaul and the processing and control burden
on the BBU pool can be greatly relieved. The edge devices (including SRRHs, F-APs and F-UEs)
of the network are assigned some new functionalities, for instance, distributed storage, control,
configuration, measurement and resource management.

As shown in Fig. 1, there are four possible transmission modes in FRAN: D2D and relay
mode, local distributed coordination mode, global centralized mode, and MRRH mode. F-UEs
will select the most appropriate mode through user-centric adaptive techniques, which take into
account the F-UEs¡¯ movement speed, communication distance, location, quality of service (QoS)
requirements, computing capability, and caching capability.

\section{Mobility Management in FRAN}
\subsection{Handover Management}
In mobile communications, handover management is one of the most critical techniques to
guarantee the QoS requirements of users. However, handover management in FRAN has not been
sufficiently studied in the existing literature. High speed F-UEs should be served by MRRHs
with large coverage areas and reliable connections. Low mobility F-UEs should be served by
SRRHs or F-APs that can provide a very high capacity to a small number of F-UEs. In a heterogeneous network, unnecessary handovers (e.g., ping-pong handovers) or handover
caused radio link failures are more likely to happen compared with conventional
cellular networks, due to the small coverage areas of small cells and severe co-channel
interference. Moreover, frequent
handovers also cause a heavy burden on the fronthaul, the backhaul and the core network. In
order to avoid frequent handovers and alleviate the control overhead of handovers, MRRHs,
SRRHs and F-APs can shift some handover related control and decision making to F-UEs in
FRAN.
\subsection{Handover Procedures in FRAN}
Since handover management in FRANs is different from that in heterogeneous networks
or in CRANs, the handover procedure should be revised for both F-AP/SRRH-to-F-AP/SRRH
handover and F-AP/SRRH-to-MRRH handover. Different from the handover procedure in CRAN,
many handover related functions such as handover and admission control are shifted from the
BBU pool to F-APs, SRRHs and MRRHs in FRAN, in conjunction with access point selection
mechanisms for various F-UEs. Moreover, in FRAN, the data traffic is generated not only in the
BBU pool but also in the network edge, such as F-UEs, F-APs and SRRHs.

Conventional handover schemes are mainly based on the received signal strength, where
handover decisions are made by comparing the received reference signal strength with a predefined
threshold. This would cause many unnecessary handovers such as ping-pong handovers in
FRANs. If access points of the mobile network are regarded as another kind of resource for
mobile devices, then the handover process can be transformed into a dynamic resource allocation
problem. Note that access points may become precious or limited resources in certain scenarios,
for example, mobile devices with urgent communication needs after an accident. It is reasonable
to prioritize the allocation of access points to such mobile devices to guarantee the QoS of the
urgent communications.
\begin{figure}[t]
        \centering
        \includegraphics*[width=13cm]{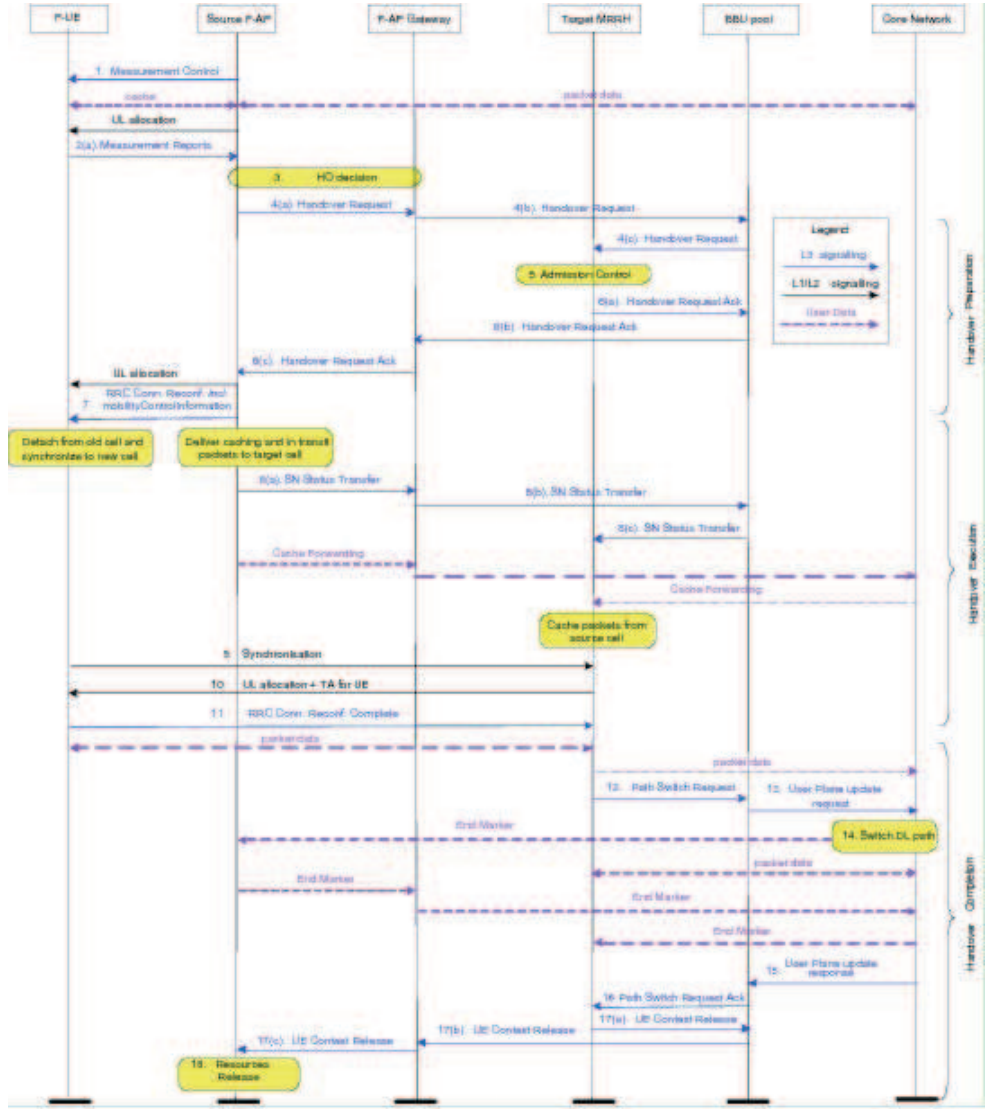}
        \caption{Handover from F-AP to MRRH.}
        \label{fig:2}
\end{figure}
\begin{figure}[t]
        \centering
        \includegraphics*[width=13cm]{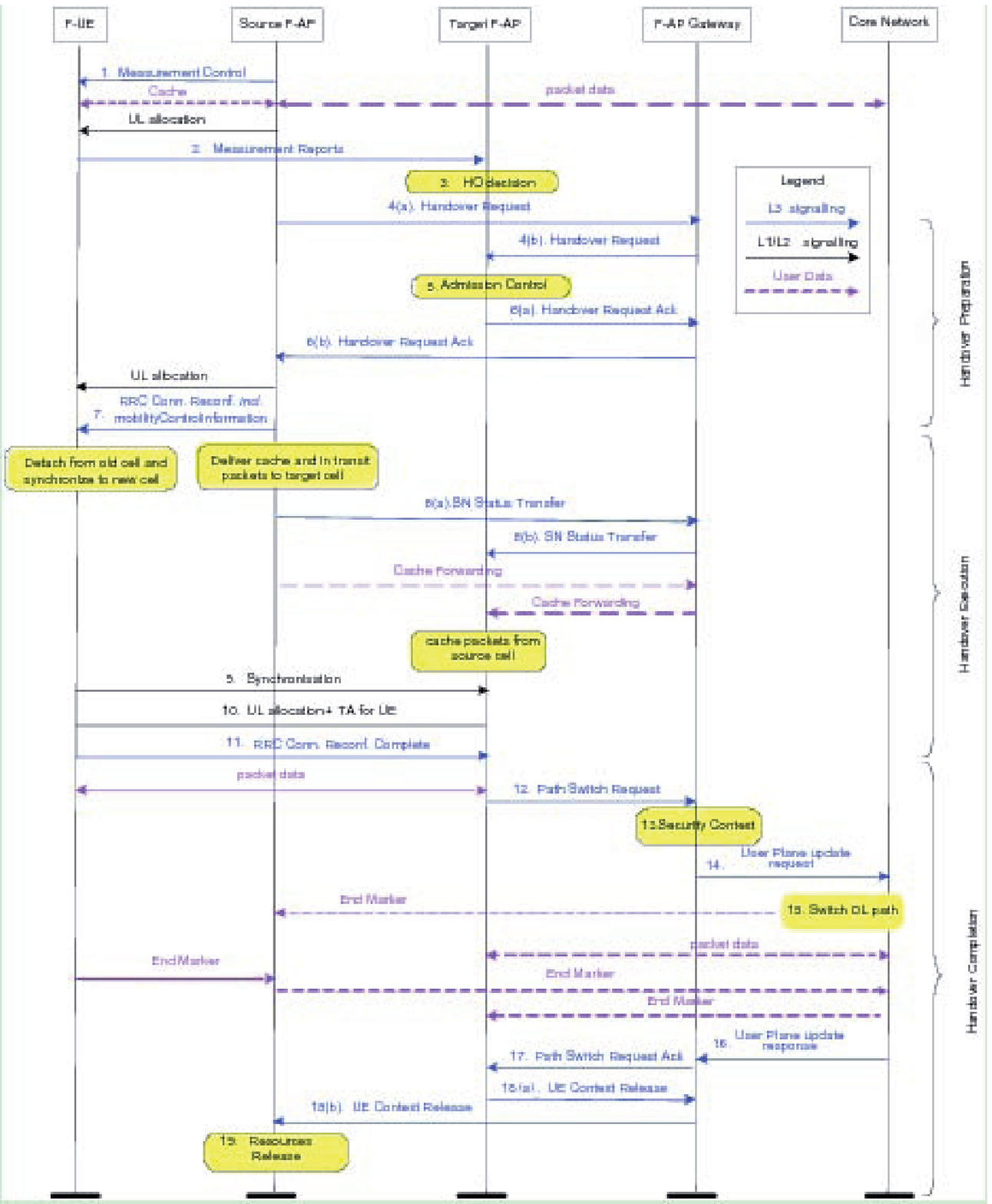}
        \caption{Handover from F-AP to F-AP.}
        \label{fig:3}
\end{figure}

Fig. 2 shows the handover procedure and signalling flow for the handover of F-UE from an F-AP
to an MRRH in FRAN. This is mainly based on the handover management in heterogeneous
cloud small cell networks \cite{Haijun2015}, but different in that the handover decisions are made between
the source F-AP and the F-AP gateway (Step 3 in Fig. 2)rather than in the BBU pool; and
the transmission of data locally cached in an F-AP to F-UE does not need to go through the
core network. Source F-AP transmits handover request signalling to target MRRH through F-AP
gateway and BBU pool (Step 4 in Fig .2). Admission control happen in target MRRH, and then
handover request Ack signalling would be transmitted to source F-AP (Step 5 in Fig .2). What
follows is the data transmission between source F-AP and target MRRH, as shown in Fig. 2. The
handover procedure in Fig. 2 can be applied to the handover from an SRRH to an MRRH as well.
The procedure of handover from an MRRH to an F-AP would be the most complex, because
there could be hundreds of possible target F-APs around and there is no direct communication about the handover from the MRRH to the F-AP.

\begin{figure}[t]
        \centering
        \includegraphics*[width=13cm]{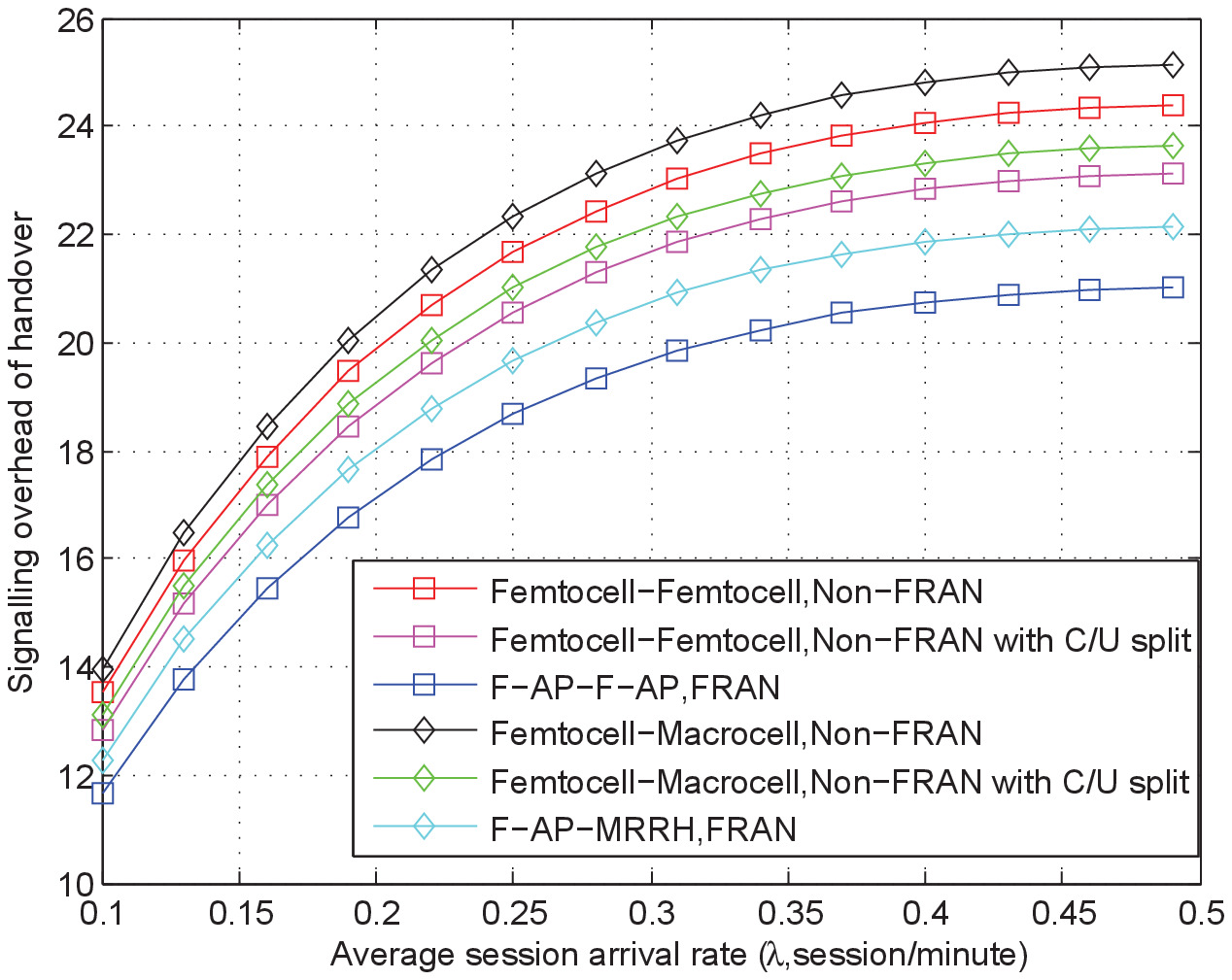}
        \caption{Signaling overhead versus average session arrival rate.}
        \label{fig:4}
\end{figure}
\begin{figure}[t]
        \centering
        \includegraphics*[width=13cm]{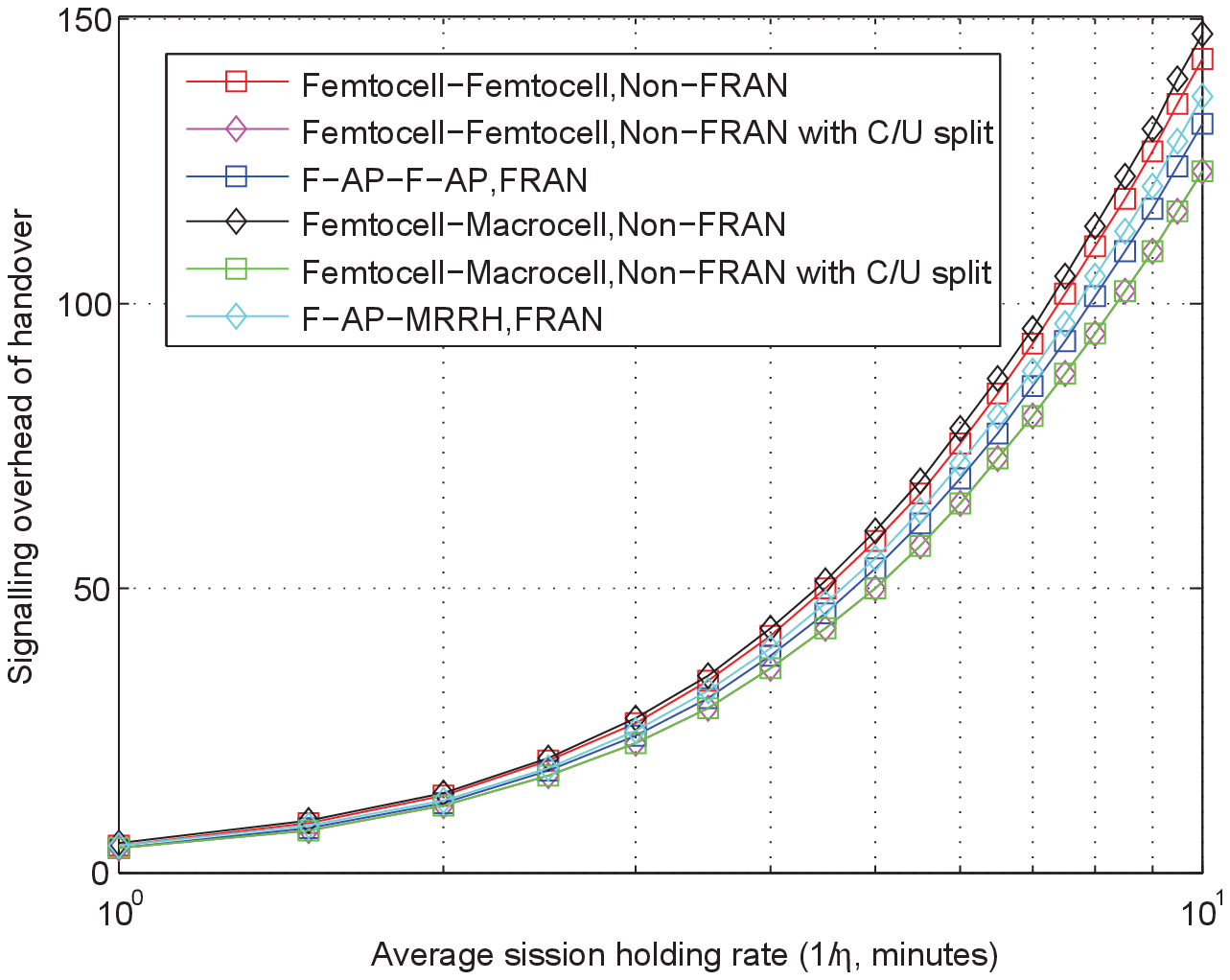}
        \caption{Signaling overhead versus average session holding rate.}
        \label{fig:5}
\end{figure}

For handover between two F-APs (or two SRRHs), the handover decisions are made between
the source F-AP (SRRH) and the F-AP (SRRH) gateway. The two F-APs (SRRHs) can communicate
with each other through the S1 interface. The handover signalling flow between the
source F-AP and the target F-AP via the F-AP gateway is shown in Fig. 3, where the FRAN
architecture deploys the traditional S1 and X2 interface. The procedure (signalling flow) of handover
between two SRRHs is similar to that in Fig. 3. Two scenarios are considered in handover between source F-AP and target F-AP/MRRH. Scenario 1: an F-UE
   in active state moves across the F-AP, whose session initializes out of F-AP, and finally moves out of the F-AP. Scenario 2: an F-UE who initializes a session under the coverage of the source F-AP remains in the active state before moving out of the source F-AP; The probability of scenario 1 ($P_{s1}$) plus the probability of scenario 2 ($P_{s2}$) equals the probability of the handover happened on the border of the F-AP/MRRH. The handover signalling overhead can be divided into processing overhead and transmitting overhead. The processing overhead consists of that at F-UE, F-AP/MRRH, F-AP gateway, and core network. The transmitting overhead includes that between F-AP and F-AP gateway, MRRH and BBU pools, and F-AP gateway and core network. Signalling overhead of the F-AP related handover in each scenario is the product of the probability of the handover in the scenario and the signalling overhead in the scenario. The signalling overhead is assumed to be proportional to the delay required to send or process a signaling message and has no unit.

We perform computer simulation to compare the performance of the proposed FRAN handover
procedure with that in a conventional RAN (denoted as non-FRAN) in terms of system signaling
overhead. In the simulation, the session holding times are generated as independent random
variables following an identical exponential distribution. The session arrivals follow a Poission
process with average arrival rate $\lambda $. The signaling overhead is assumed to be proportional to
the delay required to send or process a signaling message and has no unit. Fig. 4 shows the signaling overhead versus the average session arrival rate $\lambda $. We can see that the signaling
overhead in both FRAN and non-FRAN increases with the average session arrival rate for both
handover between F-APs and handover from an F-AP to an MRRH. This is because the number
of handovers increases with the average session arrival rate in all considered scenarios. The
non-FRAN handover procedure does not distinguish between high speed F-UEs and low speed
F-UEs. The proposed FRAN handover procedure prevents the high speed F-UEs from handing
over from MRRHs to F-APs/SRRHs, thus avoiding unnecessary handovers. Fig. 4 also shows that
for either F-AP-to-F-AP or F-AP-to-MRRH handover, the handover caused signalling overhead
in FRAN is much lower than that in non-FRAN. FRAN takes full advantage of edge devices to avoid transmitting the entire data to the BBU pool and process radio signals at the SRRHs, F-APs and F-UEs. At the same time,
the handover decisions occur between the F-AP and the F-AP gateway rather than in the BBU pool; As a result, there is the significant reduction in the transmitting overhead. And the processing overhead in F-UE and F-AP is much smaller than that in MME (mobility management entity) and core network. Thus, the handover procedure in FRAN leads to a significant reduction in signalling overhead and data traffic compared to the conventional RAN and CRAN. And the long transmission delay and heavy burden on the fronthaul (usually seen in CRAN) can be alleviated effectively.

Fig. 5 shows the signalling overhead versus the average holding time for $\lambda $ = 0.1. As the
average session holding time increases, the total signalling overhead increases. This because the
cell-boundary crossing probability increases with the session holding times, leading to a higher
probability of handover.

As shown in Fig. 4 and Fig. 5, the signaling overhead for handover in FRAN is lower than that
in non-FRAN. Moreover, as the MRRHs, SRRHs and F-APs in FRAN have control functions
and caching capabilities, a large portion of control signalling and data does not need to be
transmitted by the BBU pool. Thus, the handover procedure in FRAN leads to a significant
reduction in signalling overhead and data traffic than in conventional RAN and CRAN.

\section{Interference Management and Resource Optimization in FRAN}
Different from conventional RANs, CRANs and HCRANs, the F-APs and F-UEs in FRANs are
expected to be enhanced RRHs and UEs, respectively. F-APs integrate not only the fronthaul radio
frequency, but also local collaborative radio signal processing and cooperative radio resource management capabilities. Besides, both F-APs and F-UEs are equipped with some caching
capabilities. These distinctive characteristics of FRANs ask for a revisit of resource management
mechanism for FRANs. In an FRAN, F-UE can be collectively served by all the nearby SRRHs
and F-APs. This means that FRAN has evolved from a BS-centric architecture to a usercentric
architecture, as well as from connection-centric to content-centric. Therefore, the goal
of resource optimization in FRANs is to maximize the overall communication-plus-computing
energy efficiency, while guaranteeing the QoS requirements on transmission rates, delays and
jitters.

In FRANs, resources are not just limited to radio resources, but also include F-APs, SRRHs,
and caching and computing capabilities in F-UEs, F-APs and SRRHs. Accordingly, the resource
managment in FRANs goes beyond the traditional resource allocation to include also the allocation
of caching and computing resources at the network edge. As a promising approach to offload
traffic from the BBU pool, the D2D and relay communications enable direct communications
between mobile devices. In the following, we focus on F-APs in the discussion of resource
allocation in FRANs, while the discussion can be easily extended to SRRHs.

F-APs operate either on the same frequency band as the MRRH or on a dedicated frequency
band. On the one hand, the dedicated channel deployment of F-APs would be difficult (if not
impossible) especially when there are a large number of densely distributed F-APs. On the other
hand, cross-tier interference is serious in co-channel deployment where F-APs and MRRHs
share the same frequency band. Without effective management of cross-tier interference in the
co-channel deployment of FRAN, both the system throughput and the energy efficiency would
be largely limited. Consider a simple scenario where the FRAN contains only one MRRH and
several F-APs within the coverage area of the MRRH. The F-APs share the same frequency
band with the MRRH. Each F-AP makes decisions on the subchannel allocation and the power
allocation on each subchannel for the F-UEs associated with it.

In previous studies, game theory has been widely applied to alleviate co-channel interference
in RANs. Under the game theoretic framework, the utility function of each F-UE can be defined
as the capacity (maximum achievable data rate) of the F-UE. Accordingly, the optimization of
resource management in an FRAN can be formulated as the maximization of the overall network capacity under the constraint of maximal transmission power of F-UE. In this article, we model
the optimization of uplink subchannel and power allocation in an FRAN as a non-cooperative
game considering the selfish and rational behavior of F-UEs and F-APs. Based on the noncooperative
game, we propose an interference-aware resource (power and subchannel) allocation
scheme for the uplink of co-channel deployed F-APs. Particularly, we introduce a convex pricing
function that is proportional to the transmission power of each F-UE, in order to mitigate the
interference caused by the F-UEs to the MRRH.

The proposed interference-aware uplink resource allocation scheme starts with subchannel
allocation. The interference-aware subchannel allocation is achieved by maximizing the net utility
function of each F-UE, which is defined as the maximum achievable data rate of the F-UE
subtracted by the pricing function of the uplink interference caused by the F-UE and added
by the reward function \cite{CLF2016} of the caching offered by the F-UE. Given the optimized subchannel
allocation, the power control problem is modeled as a super-modular game. It has been proven
that the Nash Equilibriun (NE) exists on each individual subchannel \cite{Haijun2012}. When the F-UE
served by an F-AP transmits at the maximum power to achieve its maximal utility, it causes
serious uplink interference to the MRRH, thus making the strategy far from Pareto-optimality.
To approach the NE, an iterative scheme, in which the transmission power of each F-UE is
initialized at the smallest available power level and is iteratively and sequentially updated, can
be used.

We perform simulation to evaluate the performance of the proposed interference-aware uplink
resource allocation scheme. In the simulation, the spectrum sharing F-APs and the F-UEs served
by the MRRH are uniformly distributed in the central MRRH coverage area, and the F-UEs
served by F-APs are uniformly distributed in the coverage area of their serving F-AP. The
channel model of each subchannel consists of path loss and Rayleigh flat fading. In Fig. 6, we compare the total net utility function of an FRAN and that of a non-FRAN versus the number of F-UEs (femto users) per F-AP (femtocell) for three different numbers of F-APs (femtocells) in the network. The net utility gain of FRAN compared to non-FRAN performs better and better with the increasing of the number of F-UEs per F-AP because of the multi-user diversity. From Fig. 6, the performance of FRAN is still good with the increasing number of F-APs. In other words, the scheme we proposed for FRAN can provide good services, even in dense deployed networks. As shown in Fig. 6, the proposed interference-aware resource allocation scheme for FRAN performs better than the existing resource allocation for FRAN (denoted by 'Existing Scheme for FRAN' in Fig. 6) and the conventional resource allocation for LTE-A (denoted by 'non-FRAN' in Fig. 6). This is because in our proposed scheme based on the F-AP non-cooperative game, the interference pricing function in the net utility function can effectively mitigate the uplink interference from F-UEs to the MRRH, while the caching
reward function encourages F-UEs to provide local caching, thus relieving the fronthaul between
F-APs and the BBU pool and improving the total capacity of the FRAN. The total net utility
of the FRAN generally increases with the number of F-UEs per F-AP, and increases with the
number of F-APs.

\begin{figure}[t]
        \centering
        \includegraphics*[width=13cm]{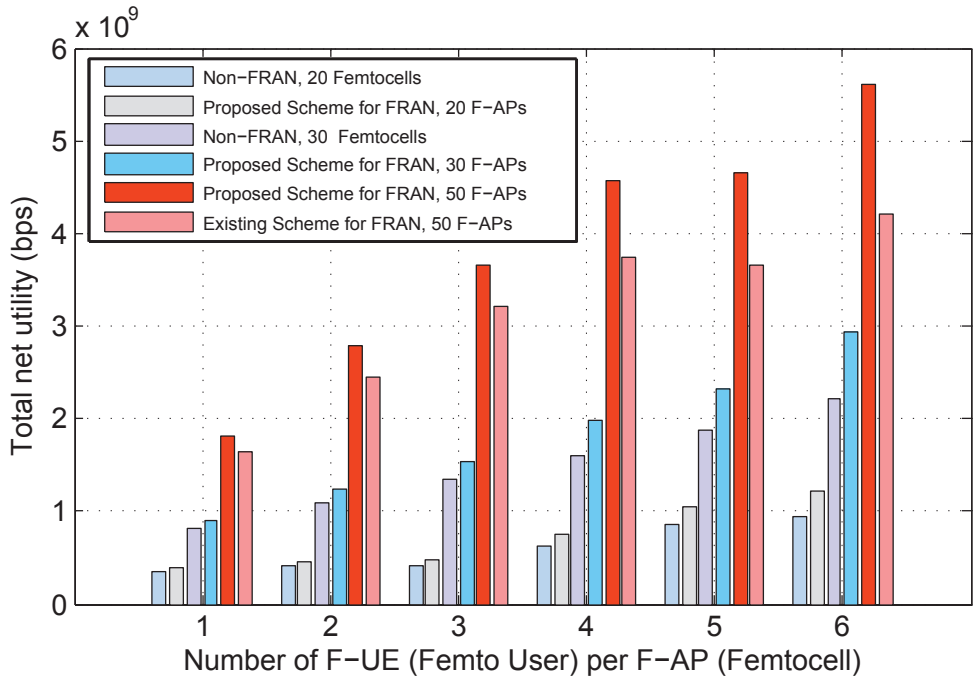}
        \caption{Total net utility of FRAN.}
        \label{fig:6}
\end{figure}

\section{Conclusion and Future Work }
In this article, we have introduced an FRAN architecture for 5G networks. The FRAN
architecture provides users with distributed local caching, computing, collaborative cooperative
radio signal processing, and cooperative resource management at the edge of the network. We
have proposed mobility management (including handover signaling procedures), interference
mitigation, and uplink resource optimization mechanisms for FRANs. Simulation results have demonstrated that the proposed FRAN architecture in conjunction with the mobility management
scheme can significantly decrease the signaling overhead of handover as compared with
conventional RANs. The proposed uplink resource allocation scheme based on an F-AP noncooperative
game can effectively mitigate the cross-tier interference and increase the total net
utility of the FRAN.

What we have discussed in this article is the portion of foundation for FRAN. There are still
many challenges and open issues that remain to be discussed in the further works. For instance,
software-defined Networking (SDN) and network function virtualization (NFV) technologies are
the most effective technique in currently practical application. How we complete the combination
of FRAN and these mature technologies is the focuses of continuing study. At the same time,
due to the distributed architecture of FRAN, there exist some security threat than CRAN, which
is of centralized systems. May be we should make the nodes in FRAN carry out selected security
functions for F-UEs. The Fog architectures should allow computing, storage, and networking tasks to be dynamically relocated among the Fog, the Cloud, and the Things. Therefore, the interfaces for Fog to interact with the Cloud, other Fogs, and the Things and users, must facilitate flexible, and in some cases dynamic, relocation of the computing, storage, and control functions among these different entities, and allow efficient and effective lifecycle management of the system and services. And Fog will provide new opportunities for us to design end-to-end systems to achieve better tradeoffs between distributed and centralized architectures, between careful deployment planning and resilience through redundancy, and between what stays local and what goes global.

Now is just the beginning, we can look forward to the changes the fog will bring to the world of networking and computing in the next decades.

\begin{IEEEbiography}{Haijun Zhang} (M'13, SM'17) is currently a Full Professor in University of Science and Technology Beijing, China. He was a Postdoctoral Research Fellow in Department of Electrical and Computer Engineering, the University of British Columbia (UBC), Vancouver Campus, Canada. Dr. Zhang has published more than 80 papers and authored 2 books. He serves as Editor of IEEE 5G Tech Focus, Journal of Network and Computer Applications, Wireless Networks, Telecommunication Systems, and KSII Transactions on Internet and Information Systems, and serves/served as a leading Guest Editor for IEEE Communications Magazine, IEEE Transactions on Emerging Topics in Computing and ACM/Springer Mobile Networks \& Applications. He serves/served as General Co-Chair of 5GWN'17 and GameNets'16, Track Chair of ScalCom2015, Symposium Chair of the GameNets'14, and Co-Chair of Workshop on 5G Ultra Dense Networks in ICC 2017, Co-Chair of Workshop on 5G Ultra Dense Networks in Globecom 2017, and Co-Chair of Workshop on LTE-U in Globecom 2017. Prof. Zhang received the IEEE ComSoc Young Author Best Paper Award in 2017.

\end{IEEEbiography}

\begin{IEEEbiography}{Yu Qiu} received the BS degree in electronic information engineering from Beijing University of Chemical Technology, Beijing, China, in 2015. She is currently pursuing the M.S. degree at the Laboratory of Wireless Communications and Networks from College of Information Science and Technology, Beijing University of Chemical Technology, Beijing, China. Her research interests include 5G networks, Fog RAN, mobility management and resource allocation in wireless communications.
\end{IEEEbiography}

\begin{IEEEbiography}{Xiaoli Chu} (S'03, M'06, SM'15) is a Senior Lecturer in the Department of Electronic and Electrical Engineering at the University of Sheffield, UK. She received the Ph.D. degree from the Hong Kong University of Science and Technology in 2005. She has co-authored more than 100 peer-reviewed journal and conference papers. She is the lead editor/author of the book Heterogeneous Cellular Networks - Theory, Simulation and Deployment, Cambridge University Press, May 2013. She is Editor for the IEEE Wireless Communications Letters and the IEEE Communications Letters. She was Co-Chair of the Wireless Communications Symposium for the IEEE International Conference on Communications 2015 (ICC'15).
\end{IEEEbiography}

\begin{IEEEbiography}{Keping Long}(SM'06) received his M.S. and Ph.D. degrees at UESTC in 1995 and 1998, respectively. He worked as an associate professor at BUPT. From July 2001 to November 2002, he was a research fellow in the ARC Special Research Centre for Ultra Broadband Information Networks (CUBIN) at the University of Melbourne, Australia. He is now a professor and dean at School of Computer and Communication Engineering (CCE), USTB. He is a member of the Editorial Committee of Sciences in China Series Fand China Communications . He is also a TPC and ISC member for COIN, IEEE IWCN, ICON, and APOC, and Organizing Co-Chair of of IWCMC'06, TPC Chair of COIN'05/'08, and TPC Co-Chair of COIN'08/'10, He was award-ed the National Science Fund Award for Distinguished Young Scholars of China in 2007 and selected as the Chang Jiang Scholars Program Professor of China in 2008. His research interests are optical Internet technology, new generation network technology, wireless information networks, value-added service, and secure network technology. He has published over 200 papers, 20 keynotes, and invited talks.
\end{IEEEbiography}

\begin{IEEEbiography}{Victor C. M. Leung} (S'75, M'89, SM'97, F'03) is a Professor of Electrical and Computer Engineering and holder of the TELUS Mobility Research Chair at the University of British Columbia. He has co-authored more than 1000 technical papers in the area of wireless networks and mobile systems. Dr. Leung is a Fellow of the Royal Society of Canada, the Canadian Academy of Engineering and the Engineering Institute of Canada. He is a winner of the 2017 Canadian Award for Telecommunications Research and the 2017 IEEE ComSoc Fred W. Ellersick Prize.
\end{IEEEbiography}


\begin{thebibliography}{99}
\bibitem{MD2011}
M. Dohler, R. W. Heath, A. Lozano, C. B. Papadias, and R. B. Valenzuela, ``Is the PHY layer dead?," \emph{IEEE Commun. Mag.}, vol. 49, no. 4, pp. 159--165, Apr. 2011.



\bibitem{AD2011}
A. Damnjanovic, J. Montojo, Y. Wei, T. Ji, T. Luo, M. Vajapeyam, T. Yoo, O. Song, and D. Malladi, ``A survey on 3GPP heterogeneous networks," \emph{IEEE Wireless Commun.}, vol. 18, no. 3, pp. 10--21, June 2011.


\bibitem{YL2010}
Y. Lin, L. Shao, Z. Zhu, Q. Wang, and R. K. Sabhikhi, ``Wireless network cloud: Architecture and system requirements," \emph{IBM J. Res. Develop.}, vol. 54, no. 1, pp. 4:1--4:12, Jan. 2010.



\bibitem{AD2013}
A. Davydov et al., ``Evaluation of joint transmission CoMP in CRAN based LTE-A HetNets with large coordination areas," \emph{IEEE GLOBECOM Wksps.}, pp. 801--806, Dec. 2013.


\bibitem{HCRANmugenIWC2014}
M. Peng, Y. Li, J. Jiang, J. Li, and C. Wang ``Heterogeneous cloud radio access networks: A new perspective for enhancing spectral and energy efficiencies," \emph{IEEE Wireless Commun.}, vol. 21, no. 6, pp. 126--135, Dec. 2014.


\bibitem{NB2014}
N. Bhushanetal , ``Network densification: The dominant theme for wireless evolution into 5G," \emph{IEEE Commun. Mag.}, vol. 52, no. 2, pp. 82--89, Feb. 2014.


\bibitem{SH2015}
S. Hung, H. Hsu, K. Chen, ``Architecture harmonization between cloud radio access networks and Fog networks," \emph{IEEE Access}, vol. 3, pp. 3019--3034, Dec. 2015.


\bibitem{DCC2014}
D. C. Chen, T. Q. S. Quek, and M. Kountouris, ``Wireless backhaul in small cell networks: Modelling and analysis,'' \emph{IEEE Veh. technol. Conf. (VTC Spring)}, May 2014.



\bibitem{FB2012}
F. Bonomi, R. Milito, J. Zhu, and S. Addepalli, ``Fog computing and its role in the Internet of things," \emph{ACM SIGCOMM}, pp. 13--16, 2012.


\bibitem{Haijun2015}
H. Zhang, C. Jiang, J. Cheng, and V. C.M. Leung , ``Cooperative interference mitigation and handover management for heterogeneous cloud small cell networks," \emph{IEEE Trans. Commun.}, vol. 22, no. 3, pp. 92--99, June 2015.


\bibitem{CLF2016}
C. Liang, F. R. Yu, H. Yao, and Z. Han, ``Virtual resource allocation in information-centric wireless networks with virtualization," \emph{IEEE Trans. Veh. Technol.}, no. 3,  Feb. 2016.


\bibitem{FZ2014}
F. Zhuang and V. Lau, ``Backhaul limited asymmetric cooperation for MIMO cellular networks via semidefinite relaxation," \emph{IEEE Trans. Signal Process.}, vol. 62, pp. 684--693, Feb. 2014.


\bibitem{RH2013}
R. Heath, S. Peters, Y. Wang, and J. Zhang, ``A current perspective on distributed antenna systems for the downlink of cellular systems,'' \emph{IEEE Commun. Mag.}, vol. 51, pp. 161--167, Apr. 2013.


\bibitem{MP2016}
M. Peng, S. Yan, K. Zhang, C. Wang, ``Fog-computing-based radio access networks: Issues and challenges," \emph{IEEE Netw.}, vol. 30, no. 4, July 2016.


\bibitem{Haijun2012}
H. Zhang, X. Chu, W. Zheng, X. Wen, ``Interference-aware resource allocation in co-channel deployment of OFDMA femtocells," \emph{IEEE Int. Conf. Commun.}, June 2012.

\end{thebibliography}
\end{document}